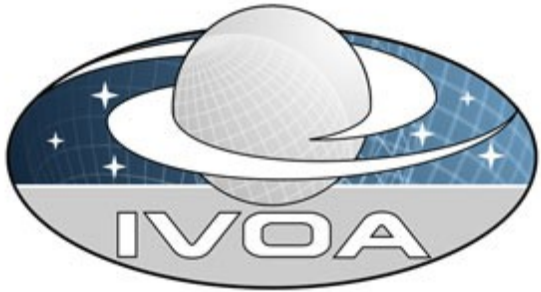

*I*nternational

*V*irtual

*O*bservatory

*A*lliance

# IVOA DataLink
# Version 1.0
# IVOA  Recommendation 2015-06-17

**Interest/Working Group:**

http://www.ivoa.net/cgi-bin/twiki/bin/view/IVOA/IvoaDAL

**This version:**

http://www.ivoa.net/documents/DataLink/20150617/

**Latest Version:**

http://www.ivoa.net/documents/DataLink/index.html

**Previous version(s):**

http://www.ivoa.net/documents/DataLink/20140413/index.html

http://www.ivoa.net/documents/DataLink/20140930/index.html

http://www.ivoa.net/documents/DataLink/20140505/index.html

http://www.ivoa.net/documents/DataLink/20140228/index.html

http://www.ivoa.net/documents/DataLink/20131022/index.html

**Editors:**

Patrick Dowler

**Authors:**

Patrick Dowler, François Bonnarel, Laurent Michel, Markus Demleitner

# Abstract

This document describes the linking of data discovery metadata to access to the data itself, further detailed metadata, related resources, and to services that perform operations on the data. The web service capability supports a drill-down into the details of a specific dataset and provides a set of links to the dataset file(s) and related resources. This specification also includes a VOTable-specific method of providing descriptions of one or more services and their input(s), usually using parameter values from elsewhere in the VOTable document. Providers are able to describe services that are relevant to the records (usually datasets with identifiers) by including service descriptors in a result document.

# Status of This Document

This is a proposed recommendation from the DAL-WG.

*This is an IVOA Proposed Recommendation made available for public review. It is appropriate to reference this document only as a recommended standard that is under review and which may be changed before it is accepted as a full recommendation.*

*A list of current IVOA Recommendations and other technical documents can be found at http://www.ivoa.net/Documents/.*

# Acknowledgments

The authors would like to thank all the participants in DAL-WG discussions for their ideas, critical reviews, and contributions to this document.

# Contents

# 1  Introduction

This specification defines mechanisms for connecting metadata about discovered datasets to the data, related data products, and web services that can act upon the data.

The *links* web service capability is a web service capability for drilling down from a discovered dataset identifier (typically an IVOA publisher dataset identifier) to find details about the data files that can be downloaded, alternate representations of the data that are available, and services that can act upon the data (usually without having to download the entire dataset). The expected usage is for DAL (Data Access Layer) data discovery services (e.g. a TAP service [6] with the ObsCore [7] data model or one of the simple DAL services) to provide an identifier that can be used to query the associated DataLink capability. The DataLink capability will respond with a list of links that can be used to access the data. Here we specify the calling interface for the capability and the response, which lists the links and provides both concrete metadata and a semantic vocabulary so clients can decide which links to use.

The *service descriptor resource* uses the metadata features of VOTable to embed service metadata along with tabular data, such as would be obtained by querying a simple DAL data discovery service or a TAP [6] service. This service metadata tells the client how to invoke a service and, for those registered in an IVOA registry, how to lookup additional information about the service. The service provider can use this mechanism to tell clients about services that can be invoked to access the discovered dataset in some way: get additional metadata, download the data, or invoke services that act upon the data files. These services may be IVOA standard services or custom services from the data providers. The current version provides no way to describe the output of a service, but this may be added in a future (minor) revision of this specification.

We expect that the *service descriptor resource* mechanism will be the primary way that clients will find and use the *links* capability from data discovery responses.

## 1.1  The Role in the IVOA Architecture

DataLink is a data access protocol in the IVOA architecture whose purpose is to provide a mechanism to link resources found via one service to resources provided by other services.

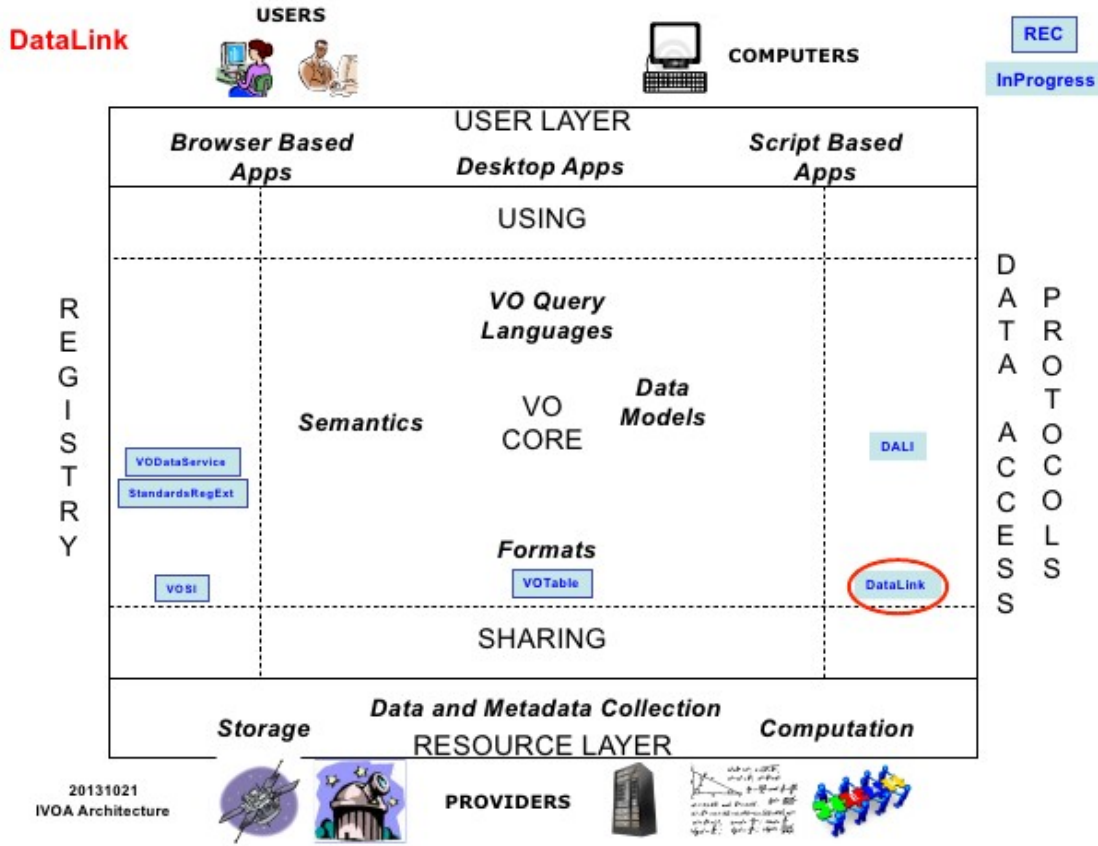

Although not shown above, any implementation of an access protocol could make use of DataLink to expose resources. DataLink services conform to the Data Access layer Interface (DALI [1]) specification, including the Virtual Observatory Support Interfaces (VOSI [2]) resources. DataLink services use VOTable [5] as the default output format for both  successful output and to return error documents.

DataLink specifies a standardID for itself, as defined in VODataService [8], to be used in a StandardsRegExt [9] record. It also specifies how to include standardID values in the response to describe links to services.

DataLink includes a description of how data discovery services can include the link to the associated DataLink service in VOTable [5]. VOTable is also the default output format for the DataLink web service capability.

## 1.2  Motivating Use Cases

Below are some of the more common use cases that have motivated the development of the DataLink specification. While this is not complete, it helps to understand the problem area covered by this specification.

### 1.2.1  Multiple Files per Dataset

It is very common for a single dataset to be physically manifest as multiple files of various types. With a DataLink web service, the client can drill down using a discovered dataset identifier and obtain links to download one or more data files. For static data files, the DataLink service will be able to provide a URL as well as the content-type and content-length (file size) for each download.

### 1.2.2  Progenitor Dataset

In some cases, the data provider may wish to provide one or more links to progenitor (input) datasets; this would enable the users to drill down to input data in order to better understand the content of the product dataset, possibly reproduce the product to evaluate the processing, or reprocess it with different parameters or software.

### 1.2.3  Alternate Representations

For some datasets (large ones) it is useful to be able to access preview data (either precomputed or generated on-the-fly) and use it to determine if the entire dataset should be downloaded (e.g. in an interactive session). A DataLink service can provide links to previews as a URL with a specific relationship to the dataset and include other metadata like content-type (e.g. image/png) and content-length to assist the client in selecting a preview; multiple previews with different sizes (content-length) could be returned in the list of links. Plots derived from the dataset could also be linked as previews. Some previews may be of the some content-type as the complete dataset, but reduced content in some fashion (e.g. a representative image or spectrum derived from a large data cube).

Links to alternate representations may be to pre-generated resources or may be computed on the fly, using either an opaque URL or a custom parameterised service (see  1.2.5  below).

Other alternate representations that are not previews could also be included in the list of links. For example, one could provide an alternate download format for a data file with different content-type (e.g. FITS and HDF).

### 1.2.4  Standard Services

Data providers often implement services that can access a dataset or its files using standard service interfaces or provide alternate representations of the dataset. For example, the links for a dataset discovered via a TAP [6] service could be to an SSA service, allowing the caller to get an SSA query response that describes the same dataset with metadata specific to the SSA service.

Providers should be able to link to current and future data access services that perform filtering and transformations as these services are defined and implemented (without requiring a new DataLink specification). For IVOA standard services, the DataLink response would use the standardID [8] as the service type to tell the client which standard (and version) the linked service complies to. The client can select services they understand and use the link to invoke the service (with additional service parameters added by the client).

### 1.2.5 Free or Custom Services

Data providers often implement custom services that can access a dataset or its files or provide alternate representations of the dataset. The availability of such services should be conveyed to clients/users in the same fashion as for standard services. This allows services defined within the VO to be used in conjunction with services defined outside the VO to deliver features to users.

### 1.2.6 Access Data Services

In many access scenarios, server-side processing of data is highly desirable, typically to reduce the amount of data to be transferred. Examples for such operations are cutouts, slicing of cubes, and re-binning to a coarser grid. Other examples for server-side operations include on-the-fly format conversion or recalibration. For the purpose of this specification, we call such services *access data* services. DataLink should let providers declare such access data services in a way that a generic client can discover what operations are supported, their semantics, and the domains of the operations' parameters. This lets clients operate multiple independent access services behind a common user interface, allowing scenarios like "give me all voxels around positions X in wavelength range Y of all spectral cubes from services Z_1, Z_2, and Z_9".

Access data services may also be standard services or custom services; at the time of writing, the definition of standard access service capabilities is in progress in separate specifications.

### 1.2.7 Recursive DataLink

In some cases, a dataset may contain many files (as in 1.2.1 above) and the provider may wish to make some files directly accessible and other (less important) files only accessible via additional calls. Such organisation of links could be accomplished by including a link to another DataLink service in the initial DataLink response (e.g. recursive DataLink). This service link would be described with both a service type (as in 1.2.4 ) and content type.

# 2 Resources

The DataLink web service capability  is implemented as an HTTP REST [11] web service capability that conforms to the DALI-sync resource [1] description. The {links} resource is described fully below (see  2.1 ). The values for the identifiers are typically found using a data discovery service (e.g. TAP with ObsCore). Data discovery services that work with an associated DataLink service should include a resource in the discovery response to describe the DataLink service. The mechanism for doing this is described below (see  4 ).

The requirements for a standalone DataLink service are given in the table below.

| resource type | resource name | required |
|---|---|---|
| {links} | service specific | yes |
| VOSI-availability | /availability | yes |
| VOSI-capabilities | /capabilities | yes |

A standalone DataLink service must have at least one {links} resource; it could have multiple {links} resources (e.g. to support alternate authentication schemes).  Alternatively, the {links} resource may be embedded in another web service, in which case the VOSI-capabilities resource of that service would also describe the {links} capability.

## 2.1  {links} resource

The {links} resource is a synchronous web service resource that conforms to the DALI-sync description [1]. The implementer is free to name for this resource however they like as long as the {links} resource is a sibling of the VOSI resources; this restriction allows a client to construct the URL to VOSI resources from any {links} URL and thus discover other capabilities or check the availablity f there is a failure. For example, a DataLink service could have:

http://example.com/datalink/links – anonymous access

http://example.com/datalink/auth-links – HTTP authentication

https://example.com/datalink/links – IVOA single-sign-on authentication

As a DALI-sync resource, the parameters for a request may be submitted using an HTTP GET (query string) or POST action.

### 2.1.1  ID

The ID parameter is used by the client to specify one or more identifiers. The service will return at least one link for each of the specified values. The ID values are found in data discovery services and may be readable URIs or opaque strings.

If an ID value specified by the client is not understandable by the DataLink service, the service must include a single link in the output with the ID and an error message (see below).

If the client submits more ID values than a service is prepared to process, the service should process ID values up to the limit and **must** include an overflow indicator in the output as described in DALI [1]. The service **must not** truncate the output within the set of rows (links) for a single ID value if the request exceeds such an input limit.

If the client submits no ID values, the service must respond with a normal response (e.g. an empty results table for VOTable output). The service may include service descriptors (see  4 ) for related services and a service descriptor describing itself (see  4.6 ).

### 2.1.2 RESPONSEFORMAT

The RESPONSEFORMAT parameter is described in DALI [1]; support for RESPONSEFORMAT is mandatory.

The only output format required by this specification is VOTable with TABLEDATA serialization [5]; services must support this format. Clients that want to get the standard (VOTable) output format should simply ignore this parameter.

To comply with this standard, a {links} resource only needs to strip off MIME type parameters and understand the following:

- no RESPONSEFORMAT

- RESPONSEFORMAT=votable

- RESPONSEFORMAT=application/x-votable+xml

All of these result in the standard output format.

Service implementers may support additional output formats but must follow the DALI [1] specification if they chose any formats described there.

## 2.2 Availability: VOSI-availability

A DataLink web service must have a VOSI-availability resource [1,2].

## 2.3 Capabilities: VOSI-capabilities

A standalone DataLink web service must have a VOSI-capabilities resource [1,2]. The standardID for the {links} resource is

```
ivo://ivoa.net/std/DataLink#links-1.0
```

The following capabilities document shows the minimal metadata and does not require a registry extension schema:

```
<?xml version="1.0" encoding="UTF-8"?>
<vosi:capabilities
   xmlns:vosi="http://www.ivoa.net/xml/VOSICapabilities/v1.0"
   xmlns:xsi="http://www.w3.org/2001/XMLSchema-instance"
   xmlns:vs="http://www.ivoa.net/xml/VODataService/v1.1">
 ...
  <capability standardID="ivo://ivoa.net/std/DataLink#links-1.0">
    <interface xsi:type="vs:ParamHTTP" role="std" version="1.0">
      <accessURL use="base">
           http://example.com/datalink/mylinks
      </accessURL>
      <queryType>GET</queryType>
      <queryType>POST</queryType>
      <resultType>
        application/x-votable+xml;content=datalink
      </resultType>
      <param std="true" use="required">
        <name>ID</name>
        <description>pubisher dataset identifier</description>
        <ucd>meta.id;meta.main</ucd>
        <dataType>string</dataType>
      </param>
    </interface>
  </capability>
</vosi:capabilities>
```

Multiple capability elements for the {links} resource may be included; this is typically used if they differ in protocol (http vs. https) and/or authentication requirements.

The {links} capability may also be included as a resource in another service, in which case the VOSI-Capabilities for that service would describe all the capabilities of that service, including {links}.

# 3  {links} Response

All responses from the {links} resource follow the rules for DALI-sync [see 1] resources, except that the {links} response allows for error messages for individual input identifier values.

## 3.1  DataLink MIME Type

In some data discovery responses (e.g. ObsCore [7]), there are columns with a URL (access_url in ObsCore) and a content type (access_format in ObsCore). If the implementation uses a DataLink service to implement this data access, it should include a complete (including the ID parameter) DataLink URL and a parameterised VOTable MIME type:

```
application/x-votable+xml;content=datalink
```

to denote that the response from that URL is a DataLink response. This is also the MIME type for the {links} response (see  3.3 ) unless the caller has explicitly requested a specific value via the RESPONSEFORMAT parameter (see  2.1.2 ). Services may include other MIME type parameters in the response.

## 3.2  List of Links

The list of links that is returned by the {links} resource can be represented as a table with the following columns:

| name | description | required | UCD |
|------|-------------|----------|-----|
| ID | Input identifier | yes | meta.id;meta.main |
| access_url | link to data or service | one only | meta.ref.url |
| service_def | reference to a service descriptor resource | | meta.ref |
| error_message | error if an access_url cannot be created | | meta.code.error |
| description | human-readable text describing this link | no | meta.note |
| semantics | Term from a controlled vocabulary describing the link | yes | meta.code |
| content_type | mime-type of the content the link returns | no | meta.code.mime |
| content_length | size of the download the link returns | no | phys.size;meta.file |

*Table 1: Required Fields for Links*

All fields must be present in the output table; values must be provided (or null) as described in the table above. Each row in the table represents one link and must have exactly one of:

- an access_url
- a service_def
- an error_message

If an error occurs while processing an ID value, there should be at least one row for that ID value and an error_message. For example, if an input ID value is not recognised or found, one row with an error_message to that effect is sufficient. If some links can be created (e.g. download links) but others cannot due to some temporary failure (e.g. service outage), then one could have one or more rows with the same ID and different error_message(s).

Services may include additional columns; this can be used to include values that can be referenced from service descriptor input parameters (see  4.1 ).

### 3.2.1  ID

The ID column contains the input identifier value.

### 3.2.2  access_url

The access_url column contains a URL to download a single resource. The URL in this column must be usable as-is with no additional parameters or client handling; it can be a link to a dynamic resource (e.g. preview generation).

### 3.2.3  service_def

The service_def column contains a reference from the result row to a separate resource. This resource describes a service as specified in section  4 . For example, if the response document includes this resource to describe a service:

```
<RESOURCE type="service" utype="adhoc:service" ID="srv1">

...

</RESOURCE>
```

then the service_def column would contain *srv1* to indicate that a resource with XML ID *srv1* in the same document describes the service. Note that service descriptors do not always require an XML ID value; it is only the reference from service_def that warrants adding an ID to the descriptor.

### 3.2.4  error_message

The error_message column is used when no accessURL can be generated for an input identifier. If an error_message is included in the output, the only other columns with values should be the ID column and the semantics column; all others should be null.

### 3.2.5  description

The description column should contain a human-readable description of the link; it is intended for display by interactive applications.

### 3.2.6  semantics

The semantics column contains a single term from an external RDF vocabulary that describes the meaning of this linked resource relative to the identified dataset. The semantics column is intended to be machine-readable and assist automating data retrieval and processing.

The core DataLink vocabulary defines a special term for the concept of *this*; this term is used to describe links used for retrieval of the dataset file(s). Since null values are not permitted, the semantics value in cases where only an error_message is supplied should be the most appropriate for the link the service was trying to generate.

The value is always interpreted as a URI; if it is a relative URI, it is resolved [11] against the base URI of the core DataLink vocabulary:

```
http://www.ivoa.net/rdf/datalink/core
```

The value used in the semantics column is normally the URI of the vocabulary, followed by a fragment (#), followed by a predicate from the specified vocabulary. For example, if the {links} table contains a link to a preview of a dataset, the ID column will contain the dataset identifier, the access_url column will contain the URL to the preview, and the semantics column could contain this predicate:

```
http://www.ivoa.net/rdf/datalink/core#preview
```

or this relative URI:

```
#preview
```

For predicates outside the core DataLink vocabulary, the full URI is required.

The core DataLink vocabulary is published at:

```
http://www.ivoa.net/rdf/datalink/core
```

and the latest version is available as a human readable document and an RDF XML document. Services are encouraged to use the core vocabulary as much as possible, but may use a custom vocabulary as long as they use a custom vocabulary namespace (base URI); the base URI should be resolvable to a human-readable document describing the terms.

### 3.2.7 content_type

The content_type column tells the client the general file format (mime-type) they will receive if they use the link (access_url or invoking a service). For recursive DataLink links, the content_type value should be as specified in section  3.1 . This field may be null (blank) if the value is unknown.

### 3.2.8 content_length

The content_length column tells the client the size of the download if they use the link, in bytes. For VOTable [5], the FIELD must be  datatype="long" with unit="byte". The value may be null (blank) if unknown and will typically be null for links to services.

## 3.3  Successful Requests

Successfully executed requests should result in a response with HTTP status code 200 (OK) and a response in the format requested by the client or in the default format for the service. The content of the response (for tabular formats) is described above, with some additional details below.

Unless the incoming request included a RESPONSEFORMAT parameter requesting a different format, the content-type header of the response MUST be application/x-votable+xml" with the "content" parameter set to "datalink", with the canonical form given in  3.1  strongly recommended.  Contrary to all other uses of the string given in  3.1 , clients wishing to evaluate the content type of the response must, however, perform a full parse of header value.  This specification cannot and does not outlaw content types with additional parameters (e.g. "application/x-votable+xml; content=datalink;charset=iso-8859-1") or with extra spaces or quotes (as allowed  for MIME types ([4]).

If the incoming request includes a DALI RESPONSEFORMAT parameter, content-type follows the DALI rules [1].

### 3.3.1  VOTable output

The table of links **must** be returned in a RESOURCE with type="results". The table **must** be in TABLEDATA serialization unless another serialization is specifically requested (see  2.1.2 ) and supported by the implementation. The name attribute for FIELD elements in the VOTable [5] (and the units in one case) are specified above (see  3.2 ).

### 3.3.2  Other Output Formats

This specification does not describe any other output formats, but allows (via the RESPONSEFORMAT in section  2.1.2 ) implementations to provide output in other formats.

## 3.4  Errors

The error handling specified for DALI-sync [see 1] resources applies to service failure (where no links can be generated) and to the usage error where no ID parameter is specified. Services should return the document format requested by the client (see  2.1.2 ). For the standard output format (VOTable) the error document **must** also be VOTable.

For errors that occur while generating individual links, each identifier may result in a link with only an error_message as described above. In either case (error document or per-link error_message), the error message must start with one of the strings in the following table, in order of specificity:

| Error | Meaning |
|---|---|
| NotFoundFault | Unknown ID value |
| UsageFault | Invalid input (e.g. invalid ID value) |
| TransientFault | Service is not currently able to function |
| FatalFault | Service cannot perform requested action |
| DefaultFault | Default failure (not covered above) |

*Table 2: Error Messages*

In all cases, the service may append additional useful information to the error strings above. If there is additional text, it must be separated from the error string with a colon (:) character, for example:

```
NotFoundFault: ivo://example.com/data?foo cannot be found
```

```
UsageFault: foo:bar is invalid, expected an ivo URI
```

# 4  Service Descriptors

The DataLink service interface is designed to add functionality to data  discovery services by providing the connection between the discovered datasets and the download of data files and access to services that act on the data. When the {links} capability returns links to services, the response document also needs to describe the services so that clients can figure out how to invoke them. This is done by including an additional metadata resource in the response document to describe each type of service that can be used.

The same mechanism can also be used in any VOTable document, such as a data discovery response from a TAP query or one of the simple DAL query protocols, to enable clients to find and use the {links} capability itself.

Here we describe how to construct a resource that describes a service and add it to a VOTable document. The mechanism is general and can be used wherever a VOTable document is created.

## 4.1  Service Resources

In a data discovery response, one RESOURCE element (usually the first) will have an attribute type="results" and tabular data; this resource contains the query result.

To describe an associated service, the VOTable would also contain one or more resources with attribute type="meta" and utype="adhoc:service". A resource of this type have no tabular data, but may include a rich set of metadata. The utype attribute makes it easy for clients to find the RESOURCE elements that describe services. A service resource contains PARAM elements to describe the service and a GROUP element with additional PARAM elements to describe the input parameters. The standard PARAM elements for a *service* resource are described in the table below.

| name | value | required |
|---|---|---|
| accessURL | URL to invoke the capability | yes |
| standardID | URI for the capability | no |
| resourceIdentifier | IVOA registry identifier | no |

*Table 3: Service Resource Parameters*

For services that implement an IVOA standard, the standardID is specified as the value attribute of the PARAM with name="standardID". For free or custom services, this PARAM is not included.

For registered services, the resourceIdentifier PARAM allows the client to query an IVOA registry for complete resource metadata. This could be used to find documentation, contact info, etc. Although they need not be, free or custom services could be registered in an IVOA registry and thus have a resourceIdentifier to enable lookup of the record.

For standard services, the value of the accessURL PARAM **must** be the accessURL for the capability specified by the standardID. The accessURL is not generally usable as-is; the client must include extra parameters as described below. If a standardID indicates a capability that supports multiple HTTP verbs (GET, POST, etc.), the client may use any supported verbs. Otherwise, there is no way in this version to specify that POST (for example) is supported so clients should assume that only HTTP GET may be used. Since the accessURL may contain parameters; clients must parse the URL to decide how to append additional parameters when invoking the service.

A GROUP with name="inputParams" contains PARAM elements describing how to invoke the service. For services where the parameter values come from columns in the results resource, we use the ref attribute of the PARAM to indicate the FIELD (column) with the values. Other PARAM elements (without a ref attribute) are also allowed; these would describe additional service parameters, the type of value that must be specified, the meaning (UCD) of the value they apply to, etc.

## 4.2 Example: Service Descriptor for the {links} Capability

The {links} capability can be used with a result table when one of the columns contains identifier values that can be used with the ID parameter (see  2.1.1 ). In order for the service resource to refer to this FIELD, the FIELD element describing this column of the table **must** include an XML ID attribute that uniquely identifies the FIELD (column). For example, a response following the ObsCore-1.0 data model would use the following:

```
<FIELD name="obs_publisher_did" ID="primaryID"
       utype="obscore:Curation.PublisherDID"
       ucd="meta.ref.url;meta.curation"
       xtype="adql:VARCHAR" datatype="char" arraysize="256*" />
```

where the ID value *primaryID* is arbitrary. This FIELD would typically be found within the RESOURCE of type="results". The same VOTable document would have a second RESOURCE with type="meta" to describe the associated DataLink {links} capability.

The {links} capability described in section  2.1  is described by following resource:

```
<RESOURCE type="meta" utype="adhoc:service">
   <PARAM name="standardID" datatype="char" arraysize="*"
        value="ivo://ivoa.net/std/DataLink#links-1.0" />
   <PARAM name="accessURL" datatype="char" arraysize="*"
        value="http://example.com/mylinks" />
   <GROUP name="inputParams">
      <PARAM name="ID" datatype="char" arraysize="*"
        value="" ref="primaryID"/>
   </GROUP>
</RESOURCE>
```

Clients that want to find services to operate on the results would look for resources with type="meta" and utype="adhoc:service". They would find a DataLink service specifically via the PARAM with name="standardID". To call the service, the  GROUP contains a PARAM with the service parameter name and a ref attribute whose value is the XML ID attribute on a FIELD. In the example above, the ref="primaryID" refers to the FIELD with ID="primaryID" in the same document (usually the result table). The URL to call the service would be:

```
http://example.com/datalink/mylinks?ID=<obs_publisher_did value>
```

Although this version of DataLink only has one parameter (ID), using a GROUP and providing the service parameter name allows this recipe to be used with any service and (with the GROUP) with multi-parameter services.

In the above example, the {links} capability is not registered in an IVOA registry so there is no resourceIdentifier PARAM included in the descriptor.

## 4.3  Example: Service Descriptor for  an SIA-1.0 Service

Suppose you have an SIA-1.0 service and you want users to be able to call it to get SIA-1.0 specific metadata. This VOTable RESOURCE describes the basic query interface of SIA-1.0:

```
<RESOURCE type="meta" utype="adhoc:service">
    <PARAM name="resourceIdentifier" datatype="char" arraysize="*"
        value="ivo://example.com/mySIA" />
    <PARAM name="standardID" datatype="char" arraysize="*"
        value="ivo://ivoa.net/std/SIA#1.0" />
    <PARAM name="accessURL" datatype="char" arraysize="*"
        value="http://example.com/sia/query" />
    <GROUP name="inputParams">
        <PARAM name="POS" datatype="char" arraysize="*"
```

```
                value=""/>
        <PARAM name="SIZE" datatype="char" arraysize="*"
            value="0.5"/>
        <PARAM name="VERB" datatype="int" value="0"/>
        <PARAM name="FORMAT" datatype="char" arraysize="*"
            value="ALL">
            <VALUES>
                <OPTION value="ALL" />
                <OPTION value="image/fits" />
                <OPTION value="METADATA" />
            </VALUES>
        </PARAM>
    </GROUP>
</RESOURCE>
```

If this SIA service supported querying specific data collections via a custom parameter named COLLECTION. The following PARAM describes the custom parameter, including the possible values:

```
<PARAM name="COLLECTION" datatype="char" arraysize="*"
            value="ALL">
        <VALUES>
          <OPTION value="ALL" />
          <OPTION value="FOO" />
          <OPTION value="BAR" />
        </VALUES>
      </PARAM>
```

This PARAM would be added to the GROUP name="input" of the service description.

## 4.4  Example: Service Descriptor for VOSpace-2.0

VOSpace-2.0 is a RESTful web service with several capabilities. Each of these capabilities can be described with a service descriptor; this would save the client having to perform a registry lookup to find and use the service. The descriptors cannot describe the path usage and XML document based input to the service, but they can describe the optional parameters:

```
<RESOURCE type="meta" utype="adhoc:service" ID="vnodes">
    <PARAM name="resourceIdentifier"
          value="ivo://example.com/vospace" />
```

```
    <PARAM name="standardID"
            value="ivo://ivoa.net/std/VOSpace/v2.0#nodes" />
    <PARAM name="accessURL"
            value="http://example.com/vospace/nodes" />

    <GROUP name="inputParams">
        <PARAM name="detail" datatype="char" arraysize="*"
                value="min"/>
        <PARAM name="limit" datatype="integer"
                value="1000"/>
        <PARAM name="uri" datatype="char" arraysize="*"
                value=""/>
    </GROUP>
</RESOURCE>
<RESOURCE type="service" utype="adhoc:service" ID="vtrans">
    <PARAM name="resourceIdentifier"
            value="ivo://example.com/vospace" />
    <PARAM name="standardID"
            value="ivo://ivoa.net/std/VOSpace/v2.0#transfers" />
    <PARAM name="accessURL"
            value="http://example.com/vospace/transfers" />
</RESOURCE>
```

Since the capability being described is RESTful and inputs are , the caller must recognise the standardID values and use a VOSpace-aware client to call the service.

## 4.5  Example: Custom Access Data Service

Parameters for custom access data services can be described such that clients can figure out how to call the service and even create a basic form-based user interface. The following Rotatable resource describes a custom spectral cutout service:

```
<RESOURCE type="meta" utype="adhoc:service" ID="apoadimo">
  <PARAM arraysize="*" datatype="char" name="accessURL"
  ucd="meta.ref.url"
  value="http://dc.zah.uni-heidelberg.de/flashheros/q/sdl/dlget"/>
```

```
  <GROUP name="inputParams">

    <PARAM arraysize="*" datatype="char" name="ID" ref="xjc7ra"
           ucd="meta.id;meta.main" value="">

      <DESCRIPTION>The pubisher DID of the dataset of
interest</DESCRIPTION>

    </PARAM>

    <PARAM arraysize="*" datatype="char" name="FLUXCALIB"
           ucd="phot.calib" utype="ssa:Char.FluxAxis.Calibration"
           value="">

      <DESCRIPTION>Recalibrate the spectrum.  Right now, the only
recalibration supported is max(flux)=1 ('RELATIVE').</DESCRIPTION>

      <VALUES>

        <OPTION name="RELATIVE" value="RELATIVE"/>

        <OPTION name="UNCALIBRATED" value="UNCALIBRATED"/>

      </VALUES>

    </PARAM>

    <PARAM ID="axi5fg" datatype="float" name="LAMBDA_MIN"
           ucd="par.min;em.wl" unit="m" value="">

      <DESCRIPTION>Spectral cutout interval, lower limit

      </DESCRIPTION>

      <VALUES>

        <MIN value="3.4211e-07"/>

        <MAX value="5.5927e-07"/>

      </VALUES>

    </PARAM>

    <PARAM ID="k4dfpe" datatype="float" name="LAMBDA_MAX"
           ucd="par.max;em.wl" unit="m" value="">

      <DESCRIPTION>Spectral cutout interval, upper limit

      </DESCRIPTION>

      <VALUES>

        <MIN value="3.4211e-07"/>

        <MAX value="5.5927e-07"/>

      </VALUES>

    </PARAM>
```

```
    <PARAM arraysize="*" datatype="char" name="FORMAT"
        ucd="meta.code.mime" utype="ssa:Access.Format" value="">
        <DESCRIPTION>MIME type of the output format</DESCRIPTION>
    <VALUES>
    <OPTION name="FITS binary table" value="application/fits"/>
    <OPTION name="Original format" value="image/fits"/>
    <OPTION name="Comma separated values" value="text/csv"/>
    <OPTION name="VOTable, tabledata encoding"
    value="application/x-votable+xml;serialization=tabledata"/>
    <OPTION name="VOTable, binary encoding"
     value="application/x-votable+xml"/>
    <OPTION name="Tab separated values"
     value="text/tab-separated-values"/>
    </VALUES>
  </PARAM>
 </GROUP>
</RESOURCE>
```

The custom service described above supports 5 input parameters: ID, FLUXCALIB, LAMBDA_MIN, LAMBDA_MAX, and FORMAT.

The PARAM describing the ID parameter has a ref attribute; the value is the XML ID of a FIELD element in a results table in the same document (the value *xjc7ra* is arbitrary; it is an opaque string that matches an ID value elsewhere in the document). The specified column contains values for the ID parameter. The client (user) will pick rows (presumably spectra) from the results table and then can invoke the service via the ID parameter and value from that row.

The FLUXCALIB parameter allows the client to specify one of two values: UNCALIBRATED or RELATIVE (listed as OPTIONS along with a description of the meaning). The UCD [10] value of phot.calib conveys the basic meaning of this parameter it is related to photometric or flux calibration).

The LAMBDA_MIN and LAMBDA_MAX parameters allow the user to specify a spectral interval to extract from the spectrum. The PARAM(s) specify that the values are wavelengths: ucd="par.min;em.wl" and ucd="par.max;em.wl" say they are minimum (par.min) and maximum (par.max) wavelength (em.wl) values. The VALUES child elements convey a range of valid wavelength values from which a subset could be extracted.

The FORMAT parameter allows the client to select from a list of output formats for the extracted spectrum. Here, the name of the PARAM is suitable to display (e.g. in a user interface) while the value would be used to call the service.

## 4.6  Example: Self-Describing Service

A service may include a service descriptor that describes itself with it's normal output. This usage is comparable to prototype work on S3 (see [12]) and when combined with calling a service with no input parameters (e.g as allowed  in  2.1.1 ) will make it easy for clients to obtain a description of both standard and custom features.

The output of a {links} capability with no input ID would include the self-describing service descriptor and an empty results table:

```
<RESOURCE type="meta" utype="adhoc:service" name="this">
   <PARAM name="standardID" datatype="char" arraysize="*"
         value="ivo://ivoa.net/std/DataLink#links-1.0" />
   <PARAM name="accessURL" datatype="char" arraysize="*"
         value="http://example.com/mylinks" />
   <GROUP name="inputParams">
      <PARAM name="ID" datatype="char" arraysize="*"
        value="" ref="primaryID"/>
   </GROUP>
</RESOURCE>
<RESOURCE type="results">
    <TABLE>
       <FIELD name="ID" datatype="char" arraysize="*"
              ucd="meta.id;meta.main" />
       <FIELD name="access_url" datatype="char" arraysize="*"
              ucd="meta.ref.url" />
       <FIELD name="service_def" datatype="char" arraysize="*"
              ucd="meta.ref" />
       <FIELD name="error_message" datatype="char" arraysize="*"
              ucd="meta.code.error" x/>
       <FIELD name="semantics" datatype="char" arraysize="*"
              ucd="meta.code" />
       <FIELD name="description" datatype="char" arraysize="*"
              ucd="meta.note" />
       <FIELD name="content_type" datatype="char" arraysize="*"
              ucd="meta.code.mime" />
       <FIELD name="content_length" datatype="long" unit="byte"
              ucd="phys.size;meta.file" />
       <DATA> <TABLEDATA/> </DATA>
```

```
    </TABLE>
</RESOURCE>
```

In the above example we give the self-describing service descriptor a name attribute with the value "this" to indicate the self-describing nature. This convention would make finding the self-description unambiguous in cases where (i) the output also contained other service descriptors and (ii) the caller could not infer which descriptor was the self-describing one from the standardID (because it is optional and not present for custom services and because they might just have a URL). Even trying to match the URL that was used with the accessURL in the descriptors is likely to be unreliable (e.g. if providers use HTTP redirects to make old URLs work when service deployment changes).

# 5 Changes

This is the initial version of this document.

## 5.1 PR-DataLink-1.0-20150413

Restricted the {links} resource path so that it must be a sibling of the VOSI resources in order to allow discovery of VOSI resources from a {links} URL.

Changed ID parameter to allow caller to invoke service with no ID values and get an empty result table; this is actually easier to implement than a special error case. Added reference to previous work on S3 and an example section where an empty links response has a self-describing service descriptor and an empty result.

Fixed URL to DALI document in the references section.

Fixed namespace prefix in example capabilities document to use recommended value.

## 5.2 PR-DataLink-1.0-20140930

Re-organised introduction to introduce the links capability and distinguish it from the service descriptor more clearly. Explicitly noted that service descriptors do not describe the output of a service.

Fixed various small typos mentioned on the RFC page.

Clarified the use of the DataLink vocabulary in the semantics column of the links table.

Changed the links table output constraints to allow only one of: access_url, service_def, or error_message. This removes the possible inconsistency of access_url in the table being different from accessURL in a service descriptor referenced by use of service_def and reduces service use by clients to a single supported approach.

Added specific datatype="long" to the content_length field in the links table.

Moved VOSpace-2.0 service descriptor to be a separate example and made it explicit that all the necessary details to invoke such a RESTful service is not supported in this version of the specification; clients must recognise the standardID to use RESTful web services.

## 5.3  PR-DataLink-20140530

Changed document status to proposed recommendation.

Removed REQUEST parameter.

Added custom service example.

Removed standard authentication and authorization error messages since these are difficult to implement consistently in different web service platforms. Changed the error message strings to use the word Fault (following GWS-WG usage, e.g. VOSpace-2.0) since Error has specific meaning in some platforms.

## 5.4  WD-DataLink-20140505

Changed the standardID for the {links} resource to include version as will be described in the StandardsRegExt record.

Changed service descriptor resource to use type="meta" utype="adhoc:service" so VOTable documents pass schema validation and this resource type can still be easily found.

Improved the VOSI-capabilities example so it describes all parameters of the example DataLink service. Added

Removed unnecessary HTTP header advice and clarified the strict DataLink mimetype usage.

Removed mention of DALI-examples since it is an optional feature for all services.

Changed name of the input parameters group element in a service descriptor to inputParams.

Fixed reference to DALI document.

Added SIA-1.0 resource desciptor example.

Tried to clarify the relationship of the two aspects of DataLink in the introduction.

Specifically allow access_url in the list of links to be different from accessURL in the service descriptor, with VOSpace example.

## 5.5 WD-DataLink-20140212

Clarified that one can implement a standalone DataLink service or include {links} resources in other services.

Re-ordered sections 2-5 so all the sections describing the {links} capability are together.

Changed from GROUP with PARAM and FIELDref siblings to PARAM with ref attribute when defining a parameter-column-with-values in section  4.1 .

Clarified the introduction so it is clear we intend to support linking of any services via RESOURCE(s) in any responses.

Changed the output of {links} resource to clearly differentiate between links with usable accessURL and links where the accessURL is a service that requires more parameters. Changed the naming style for fields in the list of links to use lower case with underscore separator so that direct potential implementations don't run into case issues.